\documentstyle[preprint,tighten,aps,epsf,floats]{revtex}

\begin{document}
\title
{Three-body recombination into deep bound states\\
in a Bose gas with large scattering length}

\author{Eric Braaten and H.-W. Hammer}

\address{Department of Physics,
 The Ohio State University, Columbus, OH 43210, USA}
\date{June 4, 2001}

\maketitle

\begin{abstract}
An effective field theory for the 
three-body system with large scattering length $a$
is applied to three-body recombination
into deep bound states in a Bose gas.
The recombination constant $\alpha$ is calculated to first
order in the short-distance interactions that allow the recombination. 
For $a<0$, the dimensionless combination
$m\alpha/(\hbar a^4)$ is a periodic function of $\ln |a|$ that
exhibits resonances 
at values of $a$ that differ by multiplicative factors of 22.7.
This dramatic behavior should be observable near a Feshbach
resonance when $a$ becomes large and negative.
\end{abstract}

\setcounter{page}{1}
\vskip 0.5cm
One important factor limiting the achievable density in Bose-Einstein 
condensates (BEC's) of trapped atoms is the loss of atoms through 3-body 
recombination \cite{SGD98,RCC00}. 
Such losses occur when three atoms scatter to form a molecular bound 
state and a third atom.
The kinetic energy of the final-state particles allows them  
to escape from the trapping potential. This
process provides a unique window on 3-body dynamics of cold atoms.
In a BEC, 3-body loss rates could also reveal
new collective mechanisms involving a molecular condensate \cite{molcond}.

The recombination event rate
can be parametrized as $\nu_{rec}=\alpha n^3$, where $\alpha$
is the recombination constant and $n$ the density of the gas.
The calculation of $\alpha$ in general is a complicated problem,
because it is sensitive to the detailed behavior of the interaction
potential. If the $s$-wave
scattering length $a$ is large compared to the size of the atoms,
however, the problem simplifies.
Assuming that $a$ is the only important length scale,
dimensional analysis implies $\alpha={\cal C} \hbar a^4/m$, 
where $m$ is the mass of the atoms and ${\cal C}$ is dimensionless.
For $a>0$, there can be 3-body recombination into a shallow $s$-wave 
state with binding energy $\hbar^2 /(ma^2)$.
For either sign of $a$, there can also be recombination into
deep molecular bound states. The coefficient $\alpha$
can then be split into two contributions $\alpha_S$ and $\alpha_D$
coming from recombination into the shallow and deep bound states,
respectively.

The 3-body recombination into the shallow bound state has been 
studied in Refs. \cite{Fed96,NiM99,EGB99,BBH00}. For the $s$-wave 
contribution, ${\cal C}$ is an oscillatory function of $\ln a$: 
\begin{equation}
\label{resalphaS}
\alpha_S \approx 67.9 \cos^2 [s_0 \ln(a\Lambda_*)+ 1.74]\,
 \frac{\hbar a^4}{m}\qquad (a>0)\,,
\end{equation}
where $s_0\approx 1.0064$ and $\Lambda_*$ is the 3-body
parameter introduced in Ref.~\cite{BBH00}.
In the adiabatic hyperspherical approximation, the oscillatory
behavior can be explained by interference between paths
connecting two hyperspherical potentials \cite{NiM99,EGB99}.
Within the framework of effective field theory, this unique behavior
is due to scaling violations created by a 3-body interaction 
required for consistent renormalization \cite{BBH00}.
For $a<0$, there is no shallow bound state (there is a low-energy virtual
state instead) and $\alpha_S \equiv 0$.

Effective field theory (EFT) is a powerful method for describing
systems composed of particles with wave number $k$ much smaller
than the inverse of the characteristic range $R$ of their interaction.
EFT focuses on the universal long-distance aspects of the problem,
by modelling the interactions as pointlike \cite{gospel}.
For wave numbers $k\ll 1/R$, one can expand
in powers of the small variable $k R$.
Generically, the scattering length $a$
is comparable to $R$, and the expansion is effectively in powers of $ka$.
In the case of large scattering length $a \gg R$,
the dependence on $ka$ must be treated nonperturbatively.
In the EFT, a single 3-body parameter is necessary and sufficient 
to renormalize the  ultraviolet divergences
at leading order in $kR$ \cite{BHK99}. The scattering length and this 3-body
parameter describe all low-energy 3-body observables
up to errors of order $R/a$.

\begin{figure}[htb]
\epsfxsize=10.cm
\centerline{\epsffile{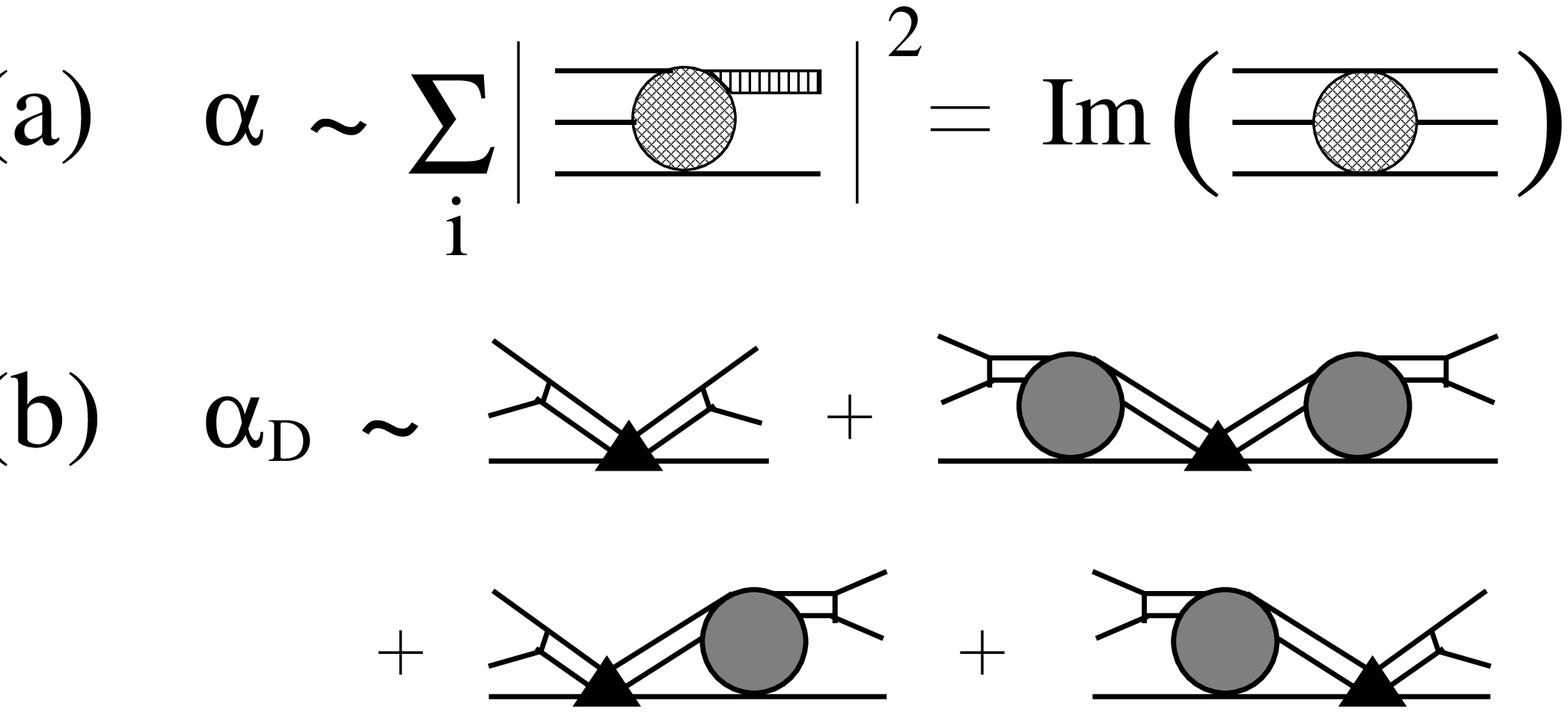}}
\vspace*{0.3cm}
\caption{(a) Illustration of the optical theorem.
Two-body bound states are indicated by the shaded double line.
(b) Diagrams for the contribution of the deep bound states.
The single (double) lines represent the exact propagators for
the field $\psi$ ($d$) and the triangle denotes an insertion of $ih'$.}
\label{fig1}
\end{figure}

In this Letter, we use EFT methods to study
the recombination constant $\alpha_D$ for deep bound states.
The deep bound states are outside the range of the effective theory 
and cannot be treated explicitly. Nevertheless, their contribution
to 3-body recombination can be calculated within
the EFT by using the optical theorem (cf.~Fig.~\ref{fig1}(a)).
The 3-body recombination channels generate
imaginary parts in the $3\to 3$ scattering amplitude. 
The imaginary part from recombination into the shallow bound state is
dynamically generated in the EFT. 
The imaginary part from 3-body recombination into deep bound states
can be taken into account by adding a local interaction term 
to the EFT. 

A similar approach has been used successfully to describe
the effects of the annihilation decays of positronium in QED and heavy
quarkonium in QCD within the framework of nonrelativistic EFT's
\cite{BBL95}. The annihilation of positronium into
photons involves intermediate states with relativistic electrons 
and positrons that are outside the range of validity of the
EFT. However, the effects of the annihilation process on 
$e^+ e^-$-scattering at low energies can be taken into 
account systematically by adding local 4-fermion operators with 
imaginary coefficients to the effective Lagrangian.
In the case of recombination into a molecule
with binding energy $B \gg \hbar^2/(m a^2)$, 
the molecule and the recoiling atom
will emerge with large momenta of order $\sqrt{mB}$
which is outside the domain of the EFT.
Up to corrections of order $p^2/(mB)$, where $p$ is the scale of 
the momentum of the three atoms, the effect of the recombination process
on the $3\to 3$ scattering amplitude can be taken into
account with a local 3-body interaction term whose coefficient
has an imaginary part. The interaction is local, 
because the three atoms have to approach to within
a distance of order $\hbar/\sqrt{mB}\ll a$ to recombine.
The imaginary 3-body coefficient
accounts collectively for recombination into all deep bound states.
It could in principle be calculated if the 2- and 3-body potentials
describing the interactions between atoms were known with sufficient
accuracy. In practice, it has to be determined from experimental data. 
Once this parameter is fixed, the dependence
of $\alpha_D$ on $a$ and $\Lambda_*$ can be predicted.

In Ref.~\cite{BBH00}, we have calculated $\alpha_S$ for a large, 
positive scattering length $a$.
In the following, we extend this work to calculate $\alpha_D$ for
large $a$ of either sign. For simplicity, we now set $\hbar=1$. 
We start with a local Lagrangian for a nonrelativistic 
boson field $\psi$ and an auxiliary field $d$ with the quantum numbers 
of two bosons \cite{BBH00,BHK99}:
\begin{eqnarray}
\label{lagt}
{\cal L}  &=&  \psi^\dagger
\bigg(i \frac{\partial}{\partial t} +\frac{\vec{\nabla}^{2}}{2m}\bigg)\psi
  + d^\dagger d \\
  &-&\frac{g}{\sqrt{2}} (d^\dagger \psi\psi +\psi^\dagger \psi^\dagger d)
   +(h+ih')\, d^\dagger d \psi^\dagger\psi +\ldots\,,\nonumber
\end{eqnarray}
\noindent
where $h'$ is the free parameter accounting for recombination
into the deep states.
The dots denote terms with more derivatives and/or fields;
those with more fields do not contribute to the 3-body amplitude,
while those with more derivatives are suppressed at low energy.
The couplings $g$ and $h$ in Eq. (\ref{lagt}) 
can be eliminated in favor of $a$ and $\Lambda_*$, respectively.
The atom propagator has the usual nonrelativistic
form $i/(\omega-p^2/2m)$. 
The 2-atom scattering amplitude for incoming atoms with 
momenta $\pm \vec{k}$ has the form $(-1/a-ik)^{-1}$. 
The exact propagator for $d$ has a pole at 
$\omega=-1/(ma^2)+\vec{p}^{\,2}/(4m)$ corresponding to the 
shallow bound state \cite{BBH00,BHK99}.

Before describing the calculation of $\alpha_D$, we
summarize the calculation of $\alpha_S$ in Ref.~\cite{BBH00}.
The coefficient for recombination into the shallow bound state can 
be expressed as
$\alpha_S= 512 \pi^2 \left|t(p_f)\right|^2 /(\sqrt{3}m)$,
where $t(p)$ is the amplitude for the transition between three atoms at rest
and a final state consisting of an atom and a shallow $s$-wave state
with momentum $p$ in their center-of-momentum frame.
This amplitude is evaluated on shell at the 
value $p_f=2/(\sqrt{3}a)$ prescribed by energy
conservation. However, $t(p)$ is also defined at off-shell values of $p$
as the solution of an integral equation that contains
the 2- and 3-body interaction terms \cite{BBH00}. 
Note that $t$ is related to the amplitude $T$ of Ref.~\cite{BBH00} via 
$t(p)=mT(p)/(48 \pi^{3/2} \sqrt{a})$.
The integral equation is regularized by a momentum cutoff $\Lambda$.
All observables can be made independent of the cutoff by adjusting the
3-body coefficient $h(\Lambda)$ as a function of $\Lambda$.
The evolution of $h$ with $\Lambda$  approaches a limit cycle:
$\Lambda^2 h$ varies periodically between $+\infty$ and $-\infty$ 
as $\Lambda$ increases by multiplicative factors of $\exp(\pi/s_0)\approx
22.7$ \cite{BHK99}. Since all observables are cutoff independent,
we can choose a $\Lambda$ for which $h(\Lambda)=0$.
The 3-body parameter $\Lambda_*$ then appears in the upper limit of the 
integral and we obtain a renormalized equation \cite{HaM00}, 
\begin{eqnarray}
\label{inteq}
& &t(p)=\frac{2}{p^2} + 
       \frac{2}{\pi}\int_0^{\Lambda_n}
      \frac{dq\,q^2\, t(q)}{-1/a+\sqrt{3}q/2-i\epsilon}\\
& &\hspace*{3cm}\times
    \,\frac{1}{pq}\ln\left|\frac{q^2+p q+p^2}{q^2-q p+p^2}\right|
    \,,\nonumber
\end{eqnarray}
with $\Lambda_n=\Lambda_* \exp[(n\pi+\arctan(1/s_0))/s_0 ]$ and
$s_0\approx 1.0064$. The integer 
$n$ should be large enough that errors of ${\cal O}(1/
(\Lambda_n a))$ can be neglected. In practice, one can often choose $n=1$.
Solving the integral equation for $t(p)$ and evaluating the solution
as described above, we obtain the result in Eq.~(\ref{resalphaS}).
The integral equation (\ref{inteq}) also has a solution for $a<0$,
but there is no shallow bound state in this case and $\alpha_S =0$.

In order to describe the recombination into
the deep bound states, we use the optical theorem as discussed
above and include an imaginary part $h'$ in the coefficient of the 
3-body contact interaction. We assume that the $h'$ term 
can be treated as a perturbation
at first order. However, the external legs have to be dressed with 
initial and final state interactions to all orders.
We have to calculate the diagrams shown in Fig.~\ref{fig1}(b),
where the blob represents the solution of Eq.~(\ref{inteq})
and the triangle represents an insertion of $ih'$. This leads to
\begin{eqnarray}
\alpha_D=\frac{288 \pi^2 a^2}{m}\frac{h_1}{\Lambda_n^2}\left|1+\frac{2}{\pi}
\int_0^{\Lambda_n}\!\!\frac{dq\,q^2\, t(q)}{-1/a+\sqrt{3}q/2-i\epsilon}
\right|^2,
\label{loop}
\end{eqnarray}
where we have defined $h'/(mg^2)=h_1/\Lambda_n^2$.
Eq.~(\ref{loop}) gives a renormalized expression for $\alpha_D$ up to 
corrections of ${\cal O}(1/(\Lambda_n a))$.
Since $t(q)$ is known numerically from solving Eq.~(\ref{inteq}),
the calculation of $\alpha_D$ entails the numerical
evaluation of the integral in Eq.~(\ref{loop}).

\begin{figure}[ht]
\epsfxsize=8.0cm
\centerline{\epsffile{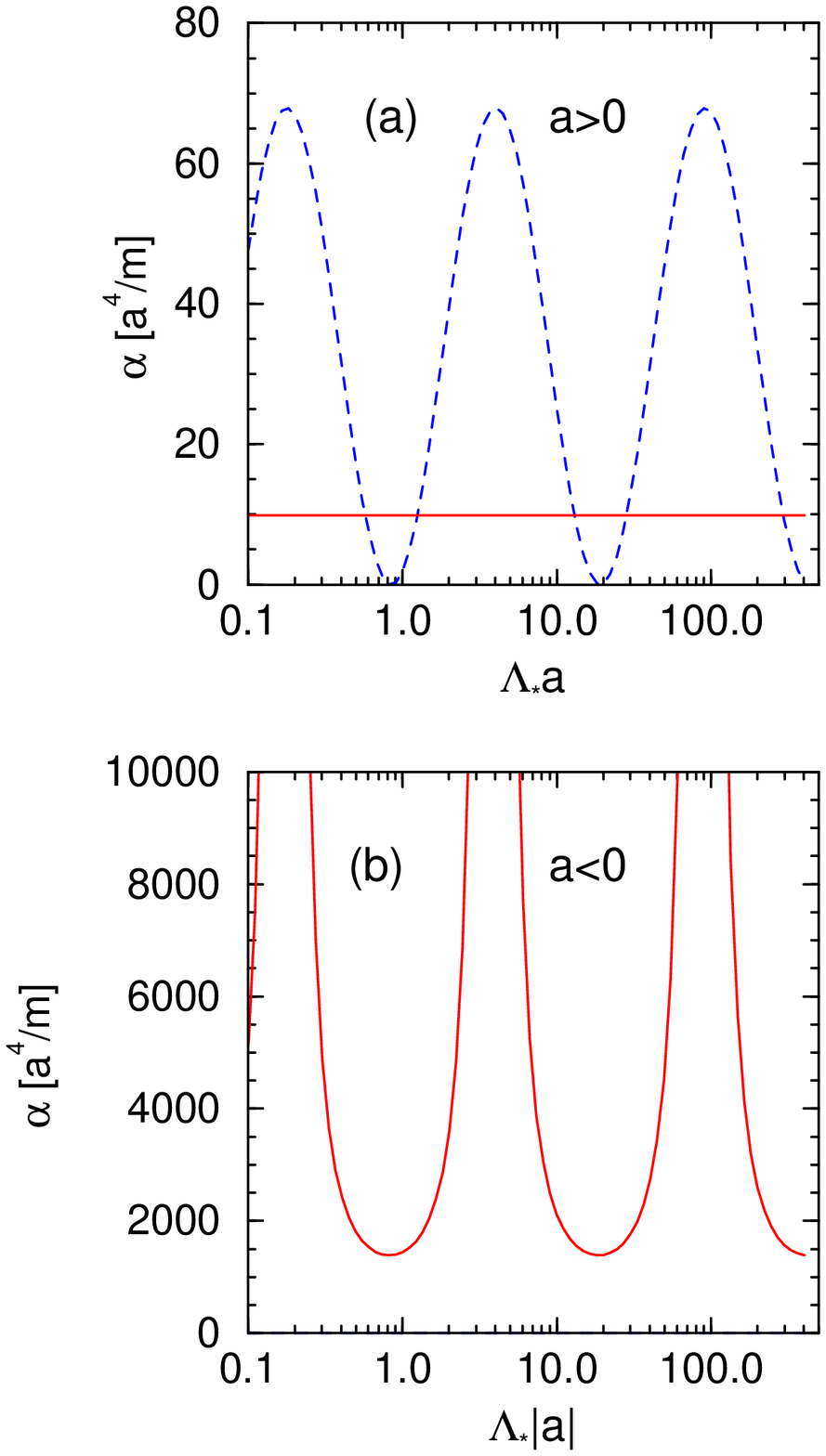}}
\vspace*{0.3cm}
\caption{The recombination coefficients
$\alpha_D$ (solid line) and  $\alpha_S$ (dashed line)
as functions of $|a|\Lambda_*$ for $a>0$ and $a<0$. The free
parameter $h_1$ in $\alpha_D$ was chosen such that
$\alpha_D=10$ for $a>0$.}
\label{fig2}
\end{figure}
There are now two cases to consider: $a>0$ and $a<0$.
We first consider the case $a>0$.
Evaluating Eq.~(\ref{loop}), we find that the amplitude is dominated by
a linear divergence in the integral and is proportional to $\Lambda_n a$.
The factors of $\Lambda_n$ cancel and the result is proportional to $a^4$
with a coefficient that, within the numerical accuracy,
is independent of $\Lambda_* a$:
\begin{equation}
\alpha_D=0.041 \,h_1\,\frac{\hbar a^4}{m}\qquad (a>0)\,.
\end{equation}
In Fig.~\ref{fig2}(a), we show both $\alpha_S$ and $\alpha_D$ for
$a > 0$, with $h_1$ adjusted such that $\alpha_D=10$.
The recombination into the deep bound states fills in the
zeroes of $\alpha_S$, so that they become at best local minima of
the recombination rate.

We now turn to the more interesting case $a<0$. Evaluating Eq.~(\ref{loop}), 
we find that the amplitude is again dominated by a linear
divergence in the integral. However, the coefficient of
$a^4/m$ is not a pure number but exhibits scaling violations.
$\alpha_D$ has the remarkable dependence
on $\Lambda_* |a|$ shown in Fig.~\ref{fig2}(b). The coefficient
diverges at a sequence of values of $\Lambda_* |a|$ that are
equally spaced on a logarithmic scale.
To a high numerical accuracy, the results can be described by the formula
\begin{equation}
\alpha_D=
\frac{5.61\,h_1}{\sin^2[s_0 \ln(\Lambda_* |a|)+1.77]}\,\frac{\hbar a^4}{m}
\qquad (a<0)\,.
\end{equation}
The oscillatory dependence on $\ln|a|$ is in accord with a general
scaling law by Efimov \cite{Efi71}.
There are divergences whenever $s_0 \ln(\Lambda_* |a|)+1.77$ is an integer
multiple of $\pi$. They occur because a 3-body Efimov state \cite{Efi71}
has been tuned to threshold by the variation of $a$.
The divergences are artifacts of the first order perturbation
theory in $h_1$, which treats the Efimov states below the 3-body
threshold as having sharp energies. However, these Efimov states 
aquire a width that depends on $h_1$ from recombination into the deep
states \cite{Estates}. This will change the divergences into resonances 
at which the recombination rate is enhanced by a factor $(\hbar^2/\Gamma 
m a^2)^2$, where $\Gamma$ is the width of an Efimov state at the
3-body threshold. First-order
perturbation theory in $h_1$ is justified if $\Gamma \ll \hbar^2/(m
a^2)$, except within $\Gamma$ of a resonance.
The determination of the height of the resonance peaks
requires a calculation that is nonperturbative in $h_1$.
Note that the elastic 3-body scattering amplitude will also
exhibit these resonances if $a<0$.
For $a>0$, an Efimov state at the 3-body threshold has a width
of order $\hbar^2/(ma^2)$ from recombination 
into the shallow 2-body bound state.
Consequently, there should be no resonances in the recombination rate.

The EFT results for $\alpha$ depend on two parameters $\Lambda_*$
and $h_1$ that are determined by interatomic interactions at distances
much smaller than $|a|$. They have predictive power only if $a$ can be
varied as a function of some external parameter and if the dependence
of $\Lambda_*$ and $h_1$ on that parameter is known. One case in which
there is predictive power is near a Feshbach resonance, where
$a$ is varied by tuning a molecular bound state close
to the threshold of two atoms by using an external magnetic field $B$.
Short-distance parameters must be smooth functions of $B$. Over the
narrow interval of a Feshbach resonance, they can be approximated
by a linear or even constant function of $B$. Linear dependence
of the short-distance parameter $g$ in (\ref{lagt}) corresponds to
$a(B)=a_0+\Delta/(B-B_0)$, where $a_0$ is the off-resonant scattering length
and $\Delta$ is the width of the Feshbach resonance.
The short distance parameters $\Lambda_*$
and $h_1$ can be approximated by constants near the resonance at $B_0$.

\begin{figure}[htb]
\epsfxsize=10.0cm
\centerline{\epsffile{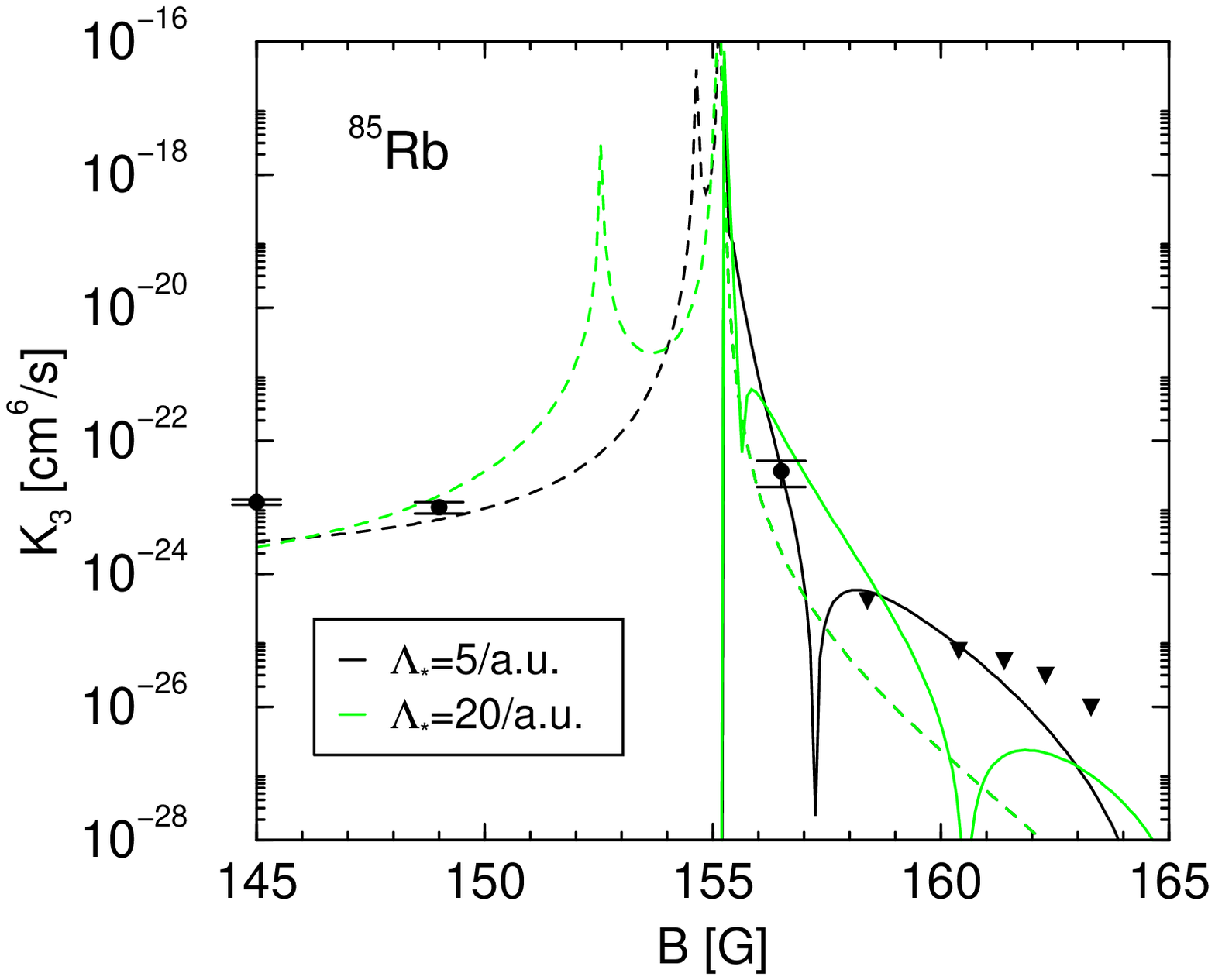}}
\vspace*{0.3cm}
\caption{The atom loss rate $K_3=3 \alpha$ for a gas of ultracold 
$^{85}$Rb-atoms. The dots (triangles) represent data (upper bounds)
for $K_3$ from Ref.~\protect\cite{RCC00}.
The EFT predictions for $K_{3S}$ (solid lines) and $K_{3D}$
(dashed lines) are shown for $\Lambda_*=5$/a.u. (black curves) and 
$\Lambda_*=20$/a.u. (grey curves). 
The arbitrary normalization factor in $K_{3D}$ is set to $h_1=24.7$.} 
\label{fig3}
\end{figure}
In Fig.~\ref{fig3} we compare our theory with measurements
of the atom loss rate coefficient $K_3=3\alpha$ in an ultracold gas of 
$^{85}$Rb-atoms by Roberts {\it et al.} \cite{RCC00}.
In $^{85}$Rb, there is a Feshbach resonance with
width $\Delta=11.5$ G at a magnetic field $B_0 =155.2$ G \cite{RCC00}.
In Fig.~\ref{fig3}, the scattering length is negative
below the Feshbach resonance at $B_0=155.2$ G and positive above.
Consequently, the shallow bound state contributes to $K_3$
only above the resonance. The free parameter $h_1=24.7$ has been fixed 
to reproduce the overall magnitude of $K_3$ below the resonance.
For $a>0$, this value of $h_1$ leads to a contribution of the deep bound 
states to the loss rate ($K_{3D}$) that is much smaller than the 
contribution of the shallow bound state ($K_{3S}$) except 
near the minima of $K_{3S}$. One might be able to
determine $h_1$ experimentally from the depth of these minima.
In Fig.~\ref{fig3}, we have plotted $K_{3D}$ and $K_{3S}$ for
two values of the 3-body parameter: 
$\Lambda_*=5/\mbox{a.u.}$ and $\Lambda_*=20/\mbox{a.u.}$.
The parameter $\Lambda_*$ would be determined very accurately if
a local minimum or maximum in the loss rate were observed.

Divergences in the 3-body recombination rate for $a<0$ have been
observed previously by Esry {\it et al.}~\cite{EGB99}. They
calculated $\alpha$ numerically for various 2-body potentials,
most of them with a single bound state (either shallow or deep). 
They varied $a$ from large negative values to large positive
values by tuning the depth of the potential. For $a>0$,
$\alpha/a^4$ increases to a maximum at some value $a_{\rm max}$
and then decreases to an interference minimum as predicted by
Eq.~(\ref{resalphaS}). For $a<0$, $\alpha/a^4$ diverges at some value
$a_{\rm div}$. The ratio $|a_{\rm div}|/a_{\rm max}$ ranged from
1.6 to 2 for the potentials studied. 
Since the maxima of $\alpha_S/a^4$ and the divergences of $\alpha_D/a^4$ 
in Fig.~\ref{fig2} appear at the same values of $|a|$, the EFT predicts
$|a_{\rm div}|/a_{\rm max}=1$. The discrepancy between the results of 
Ref.~\cite{EGB99} and the EFT prediction can be explained by a variation 
in the effective short-distance parameters $\Lambda_*$ and $h_1$
as the depth of the potential is tuned to vary $a$. Near a Feshbach
resonance, the short distance behavior of the potential is essentially 
unchanged as $B$ is varied. Thus $\Lambda_*$ and $h_1$ should be 
nearly constant and the EFT prediction should hold as long as $|a|$ is 
much larger than its off-resonance value.

We have shown that the EFT approach to the 3-body system with
large scattering length can be applied to recombination into deep
bound states. For $a>0$, this contribution fills in the interference
minima of the recombination rate into shallow bound states, transforming
them into local minima of the total recombination rate. For $a<0$,
the EFT approach makes the dramatic prediction that the recombination
rate for zero-momentum atoms should exhibit resonances due to
Efimov states at a sequence of values 
for $a$ that differ by multiplicative factors of $\exp(\pi/s_0)
\approx 22.7$.
Provided the widths of these Efimov states are numerically small
compared to $\hbar^2/(ma^2)$, the resonances should be
observable as sharp peaks in the recombination rate near a 
Feshbach resonance.

We thank C.H. Greene and J.P. Burke for useful discussions
and for providing us with numerical results of their calculations. 
This research was supported by NSF grant PHY-9800964 
and DOE grant DE-FG02-91-ER4069.


\end{document}